\documentclass[a4paper,english,aps,manuscript]{article}
\usepackage[T1]{fontenc}
\usepackage{babel}
\usepackage{graphics}

\makeatletter

\providecommand{\LyX}{L\kern-.1667em\lower.25em\hbox{Y}\kern-.125emX\@}
\newcommand{\noun}[1]{\textsc{#1}}

 \newcommand{\lyxaddress}[1]{
   \par {\raggedright #1 
   \vspace{1.4em}
   \noindent\par}
 }

\makeatother

\begin{document}

\title{Kinetic Equations for Diffusion in the Presence of Entropic Barriers}

\author{D. Reguera and J. M. Rub\'{\i} }

\maketitle

\lyxaddress{\centering Departament de F\'{\i}sica Fonamental\\
Facultat de F\'{\i}sica\\
Universitat de Barcelona\\
Diagonal 647, 08028 Barcelona, Spain\\
email: davidr@precario.ffn.ub.es }

\begin{abstract}
We use the mesoscopic nonequilibrium thermodynamics theory to derive the general
kinetic equation of a system in the presence of potential barriers. The result
is applied to the description of the evolution of systems whose dynamics is
influenced by entropic barriers. We analyze in detail the case of diffusion
in a domain of irregular geometry in which the presence of the boundaries induces
an entropy barrier when approaching the exact dynamics by a coarsening of the
description. The corresponding kinetic equation, named Fick-Jacobs equation,
is obtained, and its validity is generalized through the formulation of a scaling
law for the diffusion coefficient which depends on the shape of the boundaries.
The method we propose can be useful to analyze the dynamics of systems at the
nanoscale where the presence of entropy barriers is a common feature.
\end{abstract}

\section{Introduction}

The free energy landscape of a complex system, when represented as a function
of an order parameter or reaction coordinate, presents an intricate aspect consisting
of multiple local minima, defining metastable states, separated by barriers.
The nature of the barriers depends on which thermodynamic potential varies when
passing from one well to the other and their presence plays an important role
in the dynamics of the system.Whereas energy barriers are more frequent in problems
of solid state physics, entropy barriers are often encountered in soft condensed
matter and biological systems. These barriers may also appear when coarsening
the description of a complex system in order to simplify its dynamics. The elimination
of some coordinates can be performed by introducing a configurational entropy
accounting for the degeneracy (i.e. the number of equivalent configurations)
of the state defined by the remaining coordinates. This procedure has been used
for instance in the context of glasses \cite{stillinger}.

The study of the dynamic properties of the system at the mesoscopic level requires
the knowledge of the probability distribution function of the relevant degrees
of freedom which evolves according to a kinetic equation of the Fokker-Planck
type. This equation has been usually derived from kinetic theory in the diffusion
limit \cite{lenan}-\cite{Zubarev}, from a master equation \cite{kampen},
and from projection operator techniques \cite{kampi-oppi,deutch}.

Our purpose in this paper is to propose a new and simple theoretical framework
to analyze the mesoscopic dynamics of systems in the presence of energetic and
entropic barriers. The procedure we follow is based on the application of mesoscopic
nonequilibrium thermodynamics \cite{prigo}-\cite{ivan}, and has been previously
used to analyze transport and activated processes in systems governed by a dynamics
of the Fokker-Planck type. The kinetic equation follows after obtaining the
probability current, occurring in the corresponding continuity equation in the
configuration space, from the entropy production accounting from dissipation
in that space related to the underlying diffusion process of the probability
density. That equation applies irrespectively of the nature of the equilibrium
state of the system. When the system is in contact with a heat bath and its
volume remains unaltered, in which case the proper thermodynamic potential is
the free energy, our equation describes the dynamics in the presence of energetic
and entropic barriers. The validity of our method, however, goes beyond this
standard situation and embraces other cases of interest characterized by other
thermodynamic potentials and statistical ensembles. 

The paper is distributed in the following way. In Section II, we present a general
derivation of the kinetic equation in the framework of mesoscopic nonequilibrium
thermodynamics. That equation governs the evolution of the probability density
in the presence of barriers of any nature. In Section III we discuss an example
in which the presence of an entropic barrier manifests: diffusion in a channel
of varying cross section. The corresponding kinetic equation has been referred
to as the Fick-Jacobs equation. Section IV is devoted to the derivation of that
equation from mesoscopic nonequilibrium thermodynamics and to the formulation
of a scaling law for the diffusion coefficient which is compared to previous
results. In Section V we extend the results of Section IV to the case in which
an external field acts on the diffusing particle. Finally, in Section VI we
present our main conclusions.

\section{Kinetic equations in the presence of potential barriers}

Our purpose in this section is to present a general derivation of the kinetic
equations describing the evolution in time of the probability density when the
statics of the system is characterized by a thermodynamic potential landscape. 

We consider that the equilibrium state of the system is described by the probability
distribution function\begin{equation}
\label{peq general}
P_{eq}\sim \exp \left( \frac{-\Delta {\cal W}(\underline{x})}{kT}\right) ,
\end{equation}
where \( {\cal W}(\underline{x}) \) is the minimum reversible work required
to change the state of the system \cite{kn:landau}, which is related to the
maximum useful work which can be extracted from it, sometimes referred to as
available energy \cite{available energy} or availability \cite{availability};
\( k \) is Boltzmann's constant, \( T \) is the temperature of the medium
and \( \underline{x} \) an arbitrary set of coordinates which may represent
the velocity of a particle, the orientation of a spin, the size of a macromolecule
or whatever coordinate or order parameter whose values defines the state of
the system in a phase space. Variations of the minimum work for a thermodynamic
system are expressed as

\begin{equation}
\label{minimum work}
\Delta {\cal W}=\Delta U-T\Delta S+p\Delta V-\mu \Delta N+y\Delta Y+\ldots \, ,
\end{equation}
where \( U \) is the internal energy, \( S \) the entropy, \( V \) the volume,
and \( N \) the number of particles of the system, whereas \( T \), \( p \),
and \( \mu  \) are the temperature, pressure and chemical potential of the
environment. The term \( y\Delta Y \) represent other kinds of work (electric,
magnetic, surface work...) performed on the system, being \( y \) the intensive
parameter and \( Y \) its conjugate extensive variable \cite{reiss}. For the
case of constant temperature, volume and number of particles, the minimum work
corresponds to the Helmholtz free energy \( F \). In general, that quantity
reduces to the thermodynamics potentials by imposing to expression (\ref{minimum work})
the corresponding constraints \cite{reiss} .

Having specified the statics of the system we will derive the kinetic equation
describing the evolution of the probability density. To this purpose, we will
use the framework of mesoscopic nonequilibrium thermodynamics. This theory applies
the scheme of nonequilibrium thermodynamics to describe the dynamics of mesoscopic
degrees of freedom. The treatment of a diffusion process in the framework of
nonequilibrium thermodynamics is extended to the case in which the relevant
quantity is a probability density, defined in phase space, instead of a mass
density. The starting point is then the formulation of the Gibbs equation\begin{equation}
\label{ecuacion de Gibbs}
\delta S=-\frac{1}{T}\int \mu (\underline{x})\delta P(\underline{x},t)d\underline{x}\, ,
\end{equation}
which resembles the corresponding law proposed in nonequilibrium thermodynamics
in terms of the mass density of particles. Here \( \mu (\underline{x}) \) is
a generalized chemical potential whereas \( P(\underline{x},t) \) is the probability
density.

This expression is compatible with the Gibbs entropy

\begin{equation}
\label{postulado de entropia de Gibbs}
S(t)=-k\int P(\underline{x},t)\ln \frac{P(\underline{x},t)}{P_{eq}(\underline{x})}d\underline{x}+S_{eq}\, ,
\end{equation}
where \( S_{eq} \) is the entropy when the system and the heat bath are at
equilibrium. Effectively, taking variations in Eq. (\ref{postulado de entropia de Gibbs})
and using the expression of the chemical potential per particle \cite{degroot}

\begin{equation}
\label{potencial quimico}
\mu (\underline{x})=kT\ln \frac{P(\underline{x},t)}{P_{eq}(\underline{x})}+\mu _{eq}\, ,
\end{equation}
where \( \mu _{eq} \) is the chemical potential at equilibrium, one arrives
at expression (\ref{ecuacion de Gibbs}).

From the Gibbs equation we can obtain the entropy production related to the
underlying diffusion process in \( \underline{x} \)-space. Calculating the
time derivative of the entropy from Eq. (\ref{postulado de entropia de Gibbs})
one has

\begin{equation}
\label{variacion temporal de entropia}
\frac{\partial S}{\partial t}=-k\int \ln \frac{P}{P_{eq}}\frac{\partial P}{\partial t}d\underline{x}\, .
\end{equation}
Quite generally, we may assume that the probability density satisfies the continuity
equation 

\begin{equation}
\label{ecuacion de continuidad}
\frac{\partial P}{\partial t}=-\frac{\partial }{\partial \underline{x}}\cdot \underline{J}\, ,
\end{equation}
where \( \underline{J}(\underline{x},t) \) is a current defined in \( \underline{x} \)-space
which has to be determined. Substituting now the time derivative of Eq. (\ref{ecuacion de continuidad})
into Eq. (\ref{variacion temporal de entropia}) one obtains the entropy production

\begin{equation}
\label{produccion entropia}
\sigma =-k\int \underline{J}(\underline{x},t)\cdot \frac{\partial }{\partial \underline{x}}\ln \frac{P}{P_{eq}}\, d\underline{x}\, .
\end{equation}
To arrive at Eq. (\ref{produccion entropia}) we have also performed a partial
integration and assumed that the current vanishes at the boundaries of the system
in \( \underline{x} \)-space. 

In the nonequilibrium thermodynamics scheme, the entropy production (\ref{produccion entropia})
consists of contributions of products between the current \( \underline{J}(\underline{x},t) \)
and its conjugated thermodynamic force \( -k\frac{\partial }{\partial \underline{x}}\ln \frac{P}{P_{eq}} \),
for each value of the coordinate \( \underline{x}. \) Assuming locality in
\( \underline{x} \)-space, for which only currents and forces at the same value
of \( \underline{x} \) are coupled, one obtains the linear law

\begin{equation}
\label{ley fenomenologica flujo}
\underline{J}=-kL(\underline{x})\frac{\partial }{\partial \underline{x}}\ln \frac{P}{P_{eq}}\, ,
\end{equation}
in which the phenomenological coefficient \( L(\underline{x}) \) may in general
depend on \( \underline{x} \). 

The resulting kinetic equation then follows by substituting Eq. (\ref{ley fenomenologica flujo})
into the continuity equation (\ref{ecuacion de continuidad})

\begin{equation}
\label{F-P}
\frac{\partial P}{\partial t}=\frac{\partial }{\partial \underline{x}}\cdot \left( DP\frac{\partial }{\partial \underline{x}}\ln \frac{P}{P_{eq}}\right) ,
\end{equation}
where we have defined the diffusion coefficient 

\begin{equation}
\label{coeficiente difusion}
D(\underline{x})=\frac{kL(\underline{x})}{P}\, .
\end{equation}
This equation, which in view of Eq. (\ref{peq general}) can also be written
as

\begin{equation}
\label{F-P drif diffusion}
\frac{\partial P}{\partial t}=\frac{\partial }{\partial \underline{x}}\cdot \left( D\frac{\partial P}{\partial \underline{x}}+\frac{D}{kT}\frac{\partial \Delta {\cal W}}{\partial \underline{x}}P\right) \, ,
\end{equation}
is the Fokker-Planck equation accounting for the evolution of the probability
density in \( \underline{x} \)-space.

Under the conditions for which \( {\cal W}=F=U-TS \), this equation transforms
into the Fokker-Planck equation for a system in the presence of energy and entropy
barriers. One then obtains\begin{equation}
\label{free energy}
\frac{\partial P}{\partial t}=\frac{\partial }{\partial \underline{x}}\cdot \left( D\frac{\partial P}{\partial \underline{x}}+\frac{D}{kT}\frac{\partial \Delta U}{\partial \underline{x}}P-\frac{D}{k}\frac{\partial \Delta S}{\partial \underline{x}}P\right) \, ,
\end{equation}
where the drift term consists of contributions due to an external potential
and to variations of the entropy. When the barrier is purely energetic, only
the first contribution remains, and the previous equation reduces to the well-known
Fokker-Planck equation \begin{equation}
\label{energia}
\frac{\partial P}{\partial t}=\frac{\partial }{\partial \underline{x}}\cdot \left( D\frac{\partial P}{\partial \underline{x}}+\frac{D}{kT}\frac{\partial \Delta U}{\partial \underline{x}}P\right) \, .
\end{equation}
If the nature of the barrier is purely entropic, Eq. (\ref{free energy}) then
corresponds to the Fick-Jacobs equation

\begin{equation}
\label{entropia}
\frac{\partial P}{\partial t}=\frac{\partial }{\partial \underline{x}}\cdot \left( D\frac{\partial P}{\partial \underline{x}}-\frac{D}{k}\frac{\partial \Delta S}{\partial \underline{x}}P\right) \, ,
\end{equation}
first derived in the context of diffusion of Brownian particles in a channel
of non-constant cross section \cite{Jacob}. 

The general form of Eq. (\ref{F-P drif diffusion}), in which the equilibrium
distribution function does not need to be specified and is given in general
by Eq. (\ref{peq general}), makes that result applicable to a great diversity
of situations. For example, for a Brownian particle, for which the minimum work
is simply its kinetic energy, that equation correspond to the usual Fokker-Planck
equation. When \( \underline{x} \) is an order parameter or a reaction coordinate,
Eq. (\ref{F-P drif diffusion}) provides the kinetic equation in the presence
of barriers. The method used in this section then offers a common formalism
able to analyze the dynamics of a system in the presence of energy and entropy
barriers. In this paper, we will mainly focus our analysis to the case in which
the potential is strictly entropic.

\section{Diffusion through a channel of varying cross section: the Fick-Jacobs equation}

The influence that the presence of an entropic barrier exerts on the dynamics
of a system can be illustrated by deriving the kinetic equation governing Brownian
diffusion through a channel of varying cross-section. This equation was first
proposed by Jacobs \cite{Jacob} and subsequently rederived by Zwanzig \cite{Zwanzig Chem}
on the basis of more fundamental arguments.

\subsection{Jacobs' derivation}

Jacobs in his book \emph{Diffusion Processes} \cite{Jacob} provided an heuristic
derivation of the equation governing diffusion in a symmetric tube whose cross
section \( A(x) \) varies along the \( x \) axis, defined by the center line
of the tube. The argument runs as follows.

Consider an elementary volume of thickness \( dx \) perpendicular to the axis
of the tube. The total amount of particles in this slice, at \( x \) and time
\( t \) is \( C(x,t)dx \), which is the integral of the concentration over
the volume \( A(x)dx \). The rates of entrance and exit of the diffusing substance
into this volume are given by the Fick's law \( J=-D_{0}\frac{\partial C/A}{\partial x} \),
where \( D_{0} \) is the diffusion coefficient, and \( C/A \) is the local
volume concentration. In this case, both rates are different not only because
the concentration gradient \( \frac{\partial C}{\partial x} \) changes with
the distance, but also due to the variation of the cross-section of the channel.
Explicitly, the rate of entrance into this elementary volume is \( -D_{0}A\frac{\partial C/A}{\partial x} \)
, whereas the rate of exit is given by \( -D_{0}\left( A\frac{\partial C/A}{\partial x}+\frac{\partial }{\partial x}\left( A\frac{\partial C/A}{\partial x}\right) dx+...\right)  \).
The difference between both rates provides the rate of change of the substance
in the elementary volume, which can also be expressed as \( \frac{\partial C}{\partial t}dx \).
Neglecting quadratic terms in \( dx \), one easily arrives at the equation
governing the diffusion in the channel

\begin{equation}
\label{Fick-Jacob}
\frac{\partial C}{\partial t}=D_{0}\frac{\partial }{\partial x}\left( A\frac{\partial }{\partial x}\left( \frac{C}{A}\right) \right) =D_{0}\frac{\partial }{\partial x}\left( \frac{\partial C}{\partial x}-\frac{1}{A(x)}\frac{dA(x)}{dx}C\right) 
\end{equation}
 which is referred to as the Fick-Jacobs equation.

\subsection{Zwanzig's derivation}

Zwanzig reported in Ref. \cite{Zwanzig Chem} a more general and rigorous derivation
of the Fick-Jacobs equation. For the sake of simplicity, let us review the 2D
case. The method consists of performing a reduction in the number of coordinates
from the 2D Smoluchowski equation to a 1D description. 

The starting point is the two-dimensional Smoluchowski equation for diffusion
through a general potential \( U(x,y) \)\begin{eqnarray}
\frac{\partial c(x,y,t)}{\partial t}=D_{0}\frac{\partial }{\partial x}e^{-\beta U(x,y)}\frac{\partial }{\partial x}e^{\beta U(x,y)}c(x,y,t) &  & \\
+D_{0}\frac{\partial }{\partial y}e^{-\beta U(x,y)}\frac{\partial }{\partial y}e^{\beta U(x,y)}c(x,y,t) &  & 
\end{eqnarray}
where \( \beta =1/kT \), and \( c(x,y,t) \) is the concentration. To perform
the reduction to 1D, this equation is integrated over the variable \( y \),
leading to\begin{equation}
\label{reduccion}
\frac{\partial C(x,t)}{\partial t}=D_{0}\frac{\partial }{\partial x}\int e^{-\beta U(x,y)}\frac{\partial }{\partial x}e^{\beta U(x,y)}c(x,y,t)dy
\end{equation}
 where the reduced concentration is defined as

\begin{equation}
\label{concentracion reducida}
C(x,t)=\int c(x,y,t)dy.
\end{equation}

The key point of the derivation is the assumption of equilibration in the transverse
direction. Under this assumption, one can define an averaged \( x \)-dependent
free energy \( F(x) \) through the expression\begin{equation}
\label{x-dependent free energy}
e^{-\beta F(x)}=\int e^{-\beta U(x,y)}dy
\end{equation}
 from which one can define a normalized conditional probability distribution 

\begin{equation}
\label{probabilidad condicional}
\rho (x;y)=\frac{e^{-\beta U(x,y)}}{e^{-\beta F(x)}}.
\end{equation}
 Then, under the local equilibrium approximation, the concentration \( c(x,y,t) \)
factorizes as follows

\begin{equation}
\label{factorizacion}
c(x,y,t)\cong C(x,t)\rho (x;y)
\end{equation}

Taking these considerations into account, one finally obtains 

\begin{equation}
\label{Fick_jacob Zwanzig}
\frac{\partial C(x,t)}{\partial t}\cong \frac{\partial }{\partial x}D_{0}e^{-\beta F(x)}\frac{\partial }{\partial x}e^{\beta F(x)}C(x,t)
\end{equation}
 which constitutes the generalization of the Fick-Jacobs equation (\ref{Fick-Jacob})
to the case of a two dimensional potential \( U(x,y) \).

In the previous analysis, we have not taken into account the fact that the normal
flux must vanish at the boundaries. The role played by this zero normal flux
conditions, can be replaced by the confining potential \( U(x,y) \). In fact,
if \( U(x,y) \) is a box-like potential, i.e. zero inside the tube and infinite
outside, it is clear that \( e^{-\beta F(x)}=2w(x) \), being \( w(x) \) the
half width of the tube. In this case, the barrier is purely entropic, and one
recovers the usual Fick-Jacobs equation.

The extension to the three-dimensional case is quite straightforward by taking
into account that in that case the integration in the transverse coordinates
involves variables \( y \) and \( z \), and the width of the 2D tube must
be replaced by the transverse section \( \pi w(x)^{2} \).

Zwanzig's analysis clearly manifests that the accuracy of the Fick-Jacobs equation
is conditioned to the existence of local equilibrium in the transverse direction.
The author analyzed the effect of the deviations from the local equilibrium
related to the variations of the density \( \delta c(x,y,t)\equiv c(x,y,t)-C(x,t)\rho (x;y) \).
He derived an equation for the evolution of this deviations, which suggested
that the accuracy of the Fick-Jacobs equations is restricted to situations verifying
\( \left| w'(x)\right| <1 \), that is, when the section of the tube varies
smoothly. In addition, he showed that the range of validity of the one-dimensional
Fick-Jacobs description could be extended by introducing a position-dependent
effective diffusion coefficient \begin{equation}
\label{D zwanzig}
D_{Z}(x)=D_{0}\frac{1}{1+\gamma w'(x)^{2}}
\end{equation}
where \( D_{0} \) is the molecular diffusion coefficient and the parameter
\( \gamma  \) depends on the dimensionality. The explicit expression for this
coefficient is obtained through an expansion in powers of \( w'(x) \) \cite{Zwanzig Chem}.
The result is

\begin{equation}
\label{D expansion}
D(x)=D_{0}(1-\gamma w'(x)^{2}+...)
\end{equation}
where 

\begin{equation}
\label{exponente}
\gamma =\frac{\int \nu ^{2}e^{-\beta V(\nu )}d\nu }{\int e^{-\beta V(\nu )}d\nu },
\end{equation}
being \( \nu  \) the transverse coordinate scaled by the function \( w(x) \)
(i.e. \( \nu =\frac{y}{w(x)} \) in 2D, and \( \nu =\frac{r}{w(z)} \) for a
3D tube of radius \( r \) with cylindrical symmetry), and \( V(\nu )=U(x,y) \).
For a purely confining potential, the results obtained by Zwanzig are \( \gamma =1/3 \)
for the 2D case, and \( \gamma =1/2 \) for a 3D tube with cylindrical symmetry.
From the expansion (\ref{D expansion}) he infers the expression (\ref{D zwanzig})
based merely on the fact that this reconstruction of the series improves the
agreement with the exact results.

\section{The Fick-Jacobs equation from mesoscopic nonequilibrium thermodynamics}

For the case of an enclosure of varying cross-section, the concept of entropic
barrier is remarkably simple. At equilibrium, the density of diffusing material
\( \rho _{0} \) is constant. If we contract the 3D description retaining the
coordinate \( x \), the resulting 1D equilibrium distribution is

\begin{equation}
\label{eq en x}
\rho _{eq}(x)=\int \rho _{0}dydz=\rho _{0}A(x)\, .
\end{equation}
In this case, the diffusing particle constitute an isolated system. Therefore
its corresponding thermodynamic potential obtained form Eq. (\ref{minimum work})
is simply \( \Delta {\cal {W}}=-T\Delta S \), where the entropic barrier is,
in accordance with Eq. (\ref{peq general}), 

\begin{equation}
\label{barrera entropica}
\Delta S(x)=k\ln A(x)\, .
\end{equation}

The previous equation clearly manifests that the entropic barrier originates
from the variation of the space available for the diffusing particles. Notice
that Eq. (\ref{barrera entropica}) corresponds to the usual microcanonical
definition of the entropy in terms of the number of states which in this case
is simply proportional to the area of the tube.

Once identified the equilibrium distribution (or the entropic barrier), the
kinetic equation describing diffusion in the presence of the entropic barrier
follows from our general scheme developed in Section 2. According to Eq. (\ref{entropia}),
the kinetic equation is \begin{equation}
\label{F-J nuestra}
\frac{\partial P}{\partial t}=\frac{\partial }{\partial x}\left( D(x)\frac{\partial P}{\partial x}-\frac{D(x)}{A(x)}\frac{\partial A(x)}{\partial x}P\right) 
\end{equation}
which has the same structure as the Fick-Jacobs equation, but with a spatial
dependent diffusion coefficient.

At this point, it is worth to analyze the role played by this coefficient. In
Ref. \cite{Zwanzig Chem} it was shown that the validity of the 1D description
could be extended if we use an effective spatial-dependent diffusion coefficient.
However, the agreement one can achieve with the expression (\ref{D zwanzig})
proposed by Zwanzig is not always satisfactory. In particular, for the 3D example
discussed in Ref. \cite{Zwanzig Chem}, results for the current of particles
obtained by using the expression of the diffusion coefficient (\ref{D zwanzig})
present severe discrepancies with the corresponding expression coming from the
exact result of the 3D Smoluchowski equation, when the section of the tube changes
abruptly.

The appearance of an entropic barrier originates from the reduction of the space
to a single coordinate. But this reduction may also have implications on the
form of the diffusion coefficient. The molecular diffusion coefficient in the
real space, \( D_{0} \), gives information about the dispersion of the displacement.
In 2D, we can estimate \begin{equation}
\label{D heuristico}
D_{0}\sim \frac{(\Delta {\bf r})^{2}}{t}=\frac{\Delta x^{2}\left( 1+\left( \frac{\Delta y}{\Delta x}\right) ^{2}\right) }{t}
\end{equation}
 which manifests that the diffusion coefficient involves the displacement in
both coordinates. Once we have contracted the \( y \) coordinate, the resulting
effective diffusion coefficient only depends on the dispersion in the remaining
\( x \) coordinate, that is qualitatively \( D_{eff}\sim \frac{(\Delta x)^{2}}{t} \).
Therefore, from Eq. (\ref{D heuristico}) we can infer the behaviour \( D_{eff}\sim \frac{D_{0}}{\left( 1+\left( \frac{\Delta y}{\Delta x}\right) ^{2}\right) } \).
Previous equation provides a hint about the fact that the reduction of coordinates
may involve not only the appearance of an entropic barrier, but also a scaling
of the diffusion coefficient. Following this heuristic reasoning, we will propose
the scaling law \begin{equation}
\label{D nuestro}
D_{s}(x)=D_{0}\frac{1}{(1+y'(x)^{2})^{\alpha }}
\end{equation}
for the diffusion coefficient appearing in Eq. (\ref{F-J nuestra}), where \( y'(x)=\frac{dy}{dx} \),
and \( y(x)=w(x) \) defines the shape of the enclosure. The objective is then
to test if the use of this expression is able to extend the range of validity
of the 1D Fick-Jacob-type description. 

The values of the scaling exponents can be estimated from the calculations performed
in Ref. \cite{Zwanzig Chem}. We can expand our expression for the diffusion
coefficient \( D_{s}(x) \) in terms of \( w'(x)^{2} \) and compare the result
with the one obtained by Zwanzig (Eq. \ref{D expansion}). From that comparison,
one obtains that a reasonable choice of the scaling exponents is \( \alpha =1/3 \)
for the 2D case and \( \alpha =1/2 \) for the case of a 3D tube.

In the remaining of this section, we will test this result for the two cases
discussed in Ref. \cite{Zwanzig Chem} which admit an exact solution. We will
see that the agreement with the exact result, when we use our effective diffusion
coefficient given by Eq. (\ref{D nuestro}), is significantly improved.

\subsection{Effective Diffusion Coefficient in a 2D periodic symmetric channel }

\begin{figure}
{\par\centering \resizebox*{7.5cm}{!}{\includegraphics{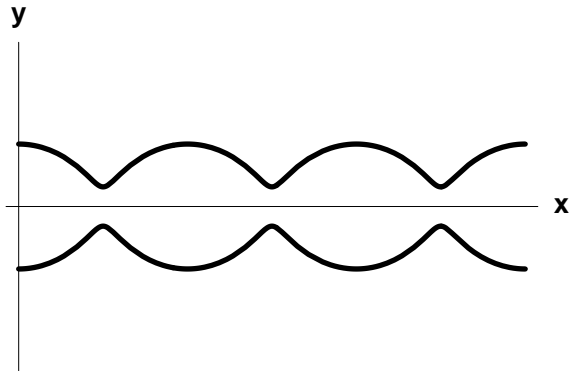}} \par}

\caption{Shape of the two-dimensional periodic channel defined parametrically by Eq.
(\ref{eq parametricas del tubo 2D}), for \protect\( V=0.5\protect \) and \protect\( a=0.5\protect \).\label{tubo 2d}}
\end{figure}

The first case under scrutiny is the diffusion in a 2D periodic channel defined
by

\begin{equation}
\label{eq parametricas del tubo 2D}
x=u+a\cosh v\sin u;\; \; \; \; \; \; \; \; \; y=v+a\sinh v\cos u
\end{equation}
where diffusion occurs in the region comprised between \( 0<u<2\pi  \) and
\( -V<v<V \). The parameter \( \lambda \equiv a\cosh V \) must be smaller
than \( 1 \), to avoid double valuation of the walls. Fig. \ref{tubo 2d} illustrates
the shape of the tube for \( V=0.5 \) and \( a=0.5 \). 

At very long times, diffusion in periodic channels can be described through
an effective diffusion coefficient \( D^{*} \), accounting for the effects
of passing through many constrictions in the channel. This effective coefficient
can be evaluated as \cite{Zwanzig PhysA}

\begin{equation}
\label{D efectivo}
\frac{1}{D^{*}}=\left\langle \frac{1}{D(x)w(x)}\right\rangle \left\langle w(x)\right\rangle 
\end{equation}
where \( \left\langle \: \right\rangle  \) denotes average over one period
of the potential.

For this particular geometry, an exact expression for \( D^{*} \) was derived
in Ref. \cite{Zwanzig PhysA} \begin{equation}
\label{D efectivo exacto}
\left( \frac{D^{*}}{D_{0}}\right) _{exact}=\left[ 1+\frac{\lambda ^{2}}{2}\frac{\tanh V}{V}\right] ^{-1}
\end{equation}
and was compared to the results obtained with the Fick-Jacobs equation using
a constant diffusion coefficient \( D_{0} \), and with the position-dependent
diffusion coefficient \( D_{Z}(x) \). 

\begin{figure}
{\par\centering \resizebox*{7.5cm}{!}{\includegraphics{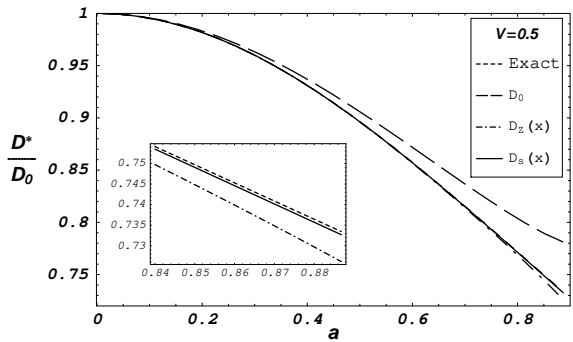}} \par}

\caption{Effective diffusion coefficient \protect\( D^{*}\protect \) for the 2D channel
for \protect\( V=0.5\protect \) as a function of the parameter \protect\( a\protect \),
using different expressions for the diffusion coefficient. The inset is a zoom
to illustrate the accuracy of the different approaches. \label{Difusion 2D}}
\end{figure}

In Fig. \ref{Difusion 2D} we have represented the results for \( D^{*} \)
obtained using the different approaches mentioned above, plus the ones corresponding
to our expression \( D_{s}(x) \). One can easily realize that the results obtained
with the expression of \( D_{s}(x) \) we propose improve the ones when using
\( D_{Z}(x) \) and constitute an excellent approximation of the exact result.

\subsection{Steady-state flux through a 3D hyperboloidal cone}

\begin{figure}
{\par\centering \resizebox*{7.5cm}{!}{\includegraphics{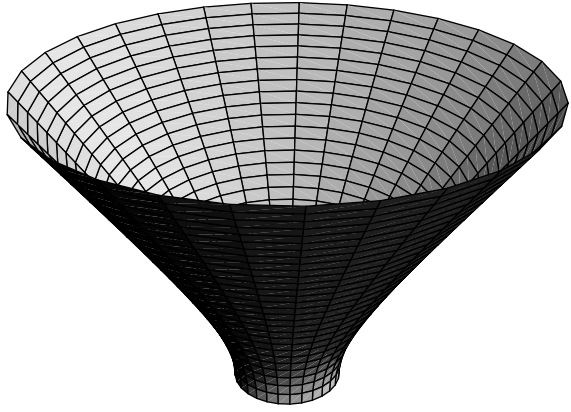}} \par}

\caption{3D hyperboloidal cone for \protect\( a=1\protect \), \protect\( 0<\xi <5\protect \)
and \protect\( \eta _{0}=0.5\protect \). \label{Tubo 3d}}
\end{figure}

The improvement of the results one can achieve with \( D_{s}(x) \) is more
remarkable for the 3D example discussed in Ref. \cite{Zwanzig Chem}. In this
case, the shape of the tube corresponds to a 3D hyperboloidal cone as the one
depicted in Fig. \ref{Tubo 3d}. It can be conveniently described by using oblate
spheroidal coordinates \( (\xi ,\eta ,\varphi ), \) related to the cylindrical
coordinates \( (r,z,\varphi ) \) by 

\begin{equation}
\label{cambio de coordenadas}
r^{2}=a^{2}(\xi ^{2}+1)(1-\eta ^{2}),\; \; \; \; \; \; z=a\xi \eta 
\end{equation}
In this coordinate system, \( \xi =0 \) corresponds to the small hole and \( \xi =\infty  \)
to the far end of the tube; \( \eta =1 \) is the z-axis, \( \eta =0 \) is
the (x,y)-plane, whereas \( \eta =\eta _{0} \) corresponds to the surface of
the tube. Diffusion then takes place in the region

\begin{equation}
\label{espacio para difusion}
0<\xi <\infty ,\; \; \; \eta _{0}<\eta <1,\; \; \; 0<\varphi <2\pi 
\end{equation}

The steady-state 3D diffusion equation can be solved by using the boundary conditions:
\( C=0 \) for \( \xi =0 \), \( C=C_{0} \) for \( \xi =\infty  \), and the
condition that the normal flux vanishes at the walls \( \eta =\eta _{0} \).
In this situation, the exact steady-state flux through the exit hole \( \xi =0 \)
is \cite{Zwanzig Chem}

\begin{equation}
\label{flujo exacto}
J_{exact}=4Da(1-\eta _{0})C_{0}
\end{equation}

\begin{figure}
{\par\centering \resizebox*{7.5cm}{!}{\includegraphics{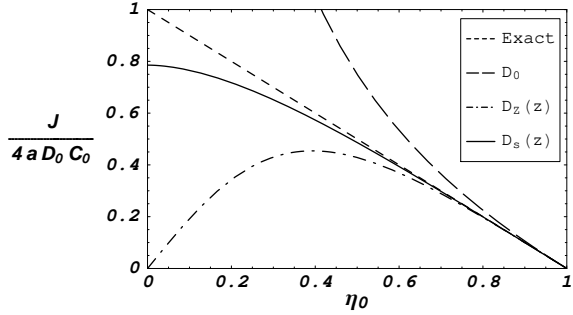}} \par}

\caption{Comparison of steady state fluxes through an arbitrary hyperboloidal cone,
against the value of \protect\( \eta _{0}\protect \).\label{Flujo 3D}}
\end{figure}

Alternatively, one can use the Fick-Jacobs equation derived in the previous
section to calculate that flux, yielding

\begin{equation}
\label{flujo de F-J}
J=C_{0}\left[ \int ^{\infty }_{0}dz\frac{1}{D(z)A(z)}\right] ^{-1}
\end{equation}
The results are depicted in Fig. \ref{Flujo 3D}, corresponding to the {}``exact{}'',
the unmodified Fick-Jacobs equation, in which \( D(z)=D_{0} \) is a constant,
and to the choices \( D_{Z}(z) \) and \( D_{s}(z) \).

From the examples we have discussed, it becomes clear that our proposal of a
scaling law for the spatial-dependent diffusion coefficient leads to a better
agreement with the exact results.

\section{Diffusion in a channel of varying cross-section under the influence of a field}

In the previous section, we have discussed the effect of entropic barriers in
the dynamics. The presence of that barriers is sometimes accompanied by energy
barriers. The theory we have developed admits a generalization to the case in
which both an entropic and an energetic barrier coexist. This will be the purpose
of this section.

\begin{figure}
{\par\centering \resizebox*{7.5cm}{!}{\includegraphics{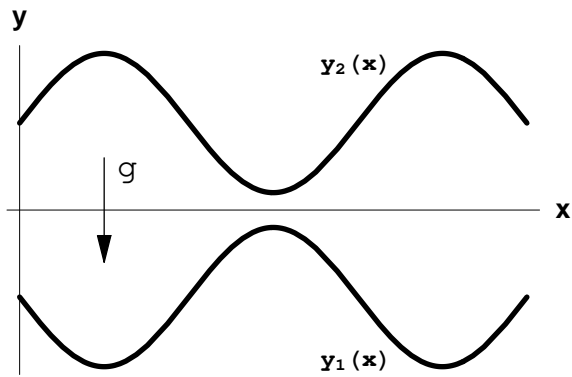}} \par}

\caption{Schematic representation of the 2D tube in the presence of gravity.\label{Gravedad}}
\end{figure}

An illustrative example of that situation is the problem of diffusion in a 2D
channel of non-constant cross-section in the presence of gravity. The situation
is depicted schematically in Fig. \ref{Gravedad}.
 If gravity is involved, the 2D equilibrium distribution is given by the usual
Boltzmann's factor\begin{equation}
\label{equilibrio gravedad}
P_{eq}(x,y)\sim e^{\frac{-mgy}{kT}}
\end{equation}
Considering the normalization condition\begin{equation}
\label{reduccion a rhox}
1=\int P_{eq}(x,y)dxdy=\int P_{eq}(x)dx
\end{equation}
and using the expression (\ref{equilibrio gravedad}), one can identify the
reduced 1D equilibrium distribution as\begin{equation}
\label{equilibrio 1D gravedad}
P_{eq}(x)\sim \int ^{y_{2}(x)}_{y_{1}(x)}dy\exp \left( \frac{-mgy}{kT}\right) =\frac{kT}{mg}\exp \left( \frac{-mgy_{1}(x)}{kT}\right) \left( 1-\exp \left( \frac{-2mgw(x)}{kT}\right) \right) 
\end{equation}
where \( y_{1}(x) \) and \( y_{2}(x) \) define the boundaries of the channel
and \( 2w(x)=y_{2}(x)-y_{1}(x) \) is its width. The dynamic equation then follows
after introducing \( P_{eq}(x) \) in Eq. (\ref{F-P}), yielding \begin{equation}
\label{Difusion gravedad tubo simetrico}
\frac{\partial P}{\partial t}=\frac{\partial }{\partial x}\left( D(x)\frac{\partial P}{\partial x}+\frac{D(x)mg}{kT}y_{1}'(x)\coth \left( \frac{mgw(x)}{kT}\right) P\right) 
\end{equation}
for the case of a symmetric channel for which \( y_{2}(x)=-y_{1}(x) \).

From the previous expressions one can distinguish different limiting cases:

\begin{enumerate}
\item In the limit \( \epsilon \equiv \frac{mgw(x)}{kT}\ll 1 \), we can expand the
reduced equilibrium distribution (\ref{equilibrio 1D gravedad}) in terms of
\( \epsilon  \) \begin{equation}
\label{expansion altas T}
P_{eq}(x)\sim \frac{kT}{mg}(1+O(\epsilon ))\left( 1-1+\frac{2mgw(x)}{kT}+O(\epsilon ^{2})\right) \sim 2w(x),
\end{equation}
 recovering the case of purely entropic barrier discussed in the previous section.
Eq. (\ref{Difusion gravedad tubo simetrico}) reduces then to the Fick-Jacobs
equation (\ref{F-J nuestra}).
\item For the case \( \epsilon \gg 1 \), gravity dominates, \( \coth \epsilon \rightarrow 1 \),
and we recover the dynamics for diffusion along a 1D purely energetic barrier.\begin{equation}
\label{Difusion con gravedad sin confinar}
\frac{\partial P}{\partial t}=\frac{\partial }{\partial x}\left( D(x)\frac{\partial P}{\partial x}+\frac{D(x)mg}{kT}y_{1}'(x)P\right) 
\end{equation}

\item If the boundaries are flat, that is \( y_{1}(x)=-y_{2}(x)=cte \), the equilibrium
distribution \( P_{eq}(x) \) is then constant, and the dynamics can be properly
described by the 1D Smoluchowski equation.\begin{equation}
\label{Fick}
\frac{\partial P}{\partial t}=\frac{\partial }{\partial x}\left( D(x)\frac{\partial P}{\partial x}\right) 
\end{equation}

\end{enumerate}
The evolution of this confined system in the presence of gravity dictated by
Eq. (\ref{Difusion gravedad tubo simetrico}) presents some peculiarities. On
one hand, since the energy barrier depends on the coordinate we have eliminated,
there exists a coupling between entropy and energy barriers and not a mere superposition
of the drifts related to each of them, which is reflected by the presence of
the term \( \coth \left( \frac{mgw(x)}{kT}\right)  \) in the drift of Eq. (\ref{Difusion gravedad tubo simetrico}).
On the other hand, it is important to highlight that Eq. (\ref{Difusion gravedad tubo simetrico})
does not satisfy the detailed balance condition: the mobility \( b\equiv \frac{D(x)}{kT}\coth \left( \frac{mgw(x)}{kT}\right)  \)
and the diffusion \( D(x) \) are not related through the usual Einstein relation
\( D(x)=kTb \). That is the signature that the fluctuation-dissipation theorem,
which holds at equilibrium, may lose its validity when we perform a reduction
of the variables describing the state of the system. 

Concerning the effective diffusion coefficient, a scaling law as the one proposed
through Eq. (\ref{D nuestro}), which is valid in absence of external forces,
will be in general not correct when an external field is present. The value
of the scaling exponent \( \alpha  \) varies in the presence of external potentials,
since the scaling exponents changes due to interactions. The range of validity
of the Fick-Jacobs description and the modifications in the scaling of the effective
diffusion coefficient under the presence of external fields requires a more
elaborated treatment.

\section{Conclusions}

In this paper we have presented a theory to describe the kinetics of a system
whose equilibrium state is characterized by a given landscape of an unspecified
thermodynamic potential. The theory is based on mesoscopic non-equilibrium thermodynamics,
which uses the scheme of nonequilibrium thermodynamics \cite{degroot} at the
mesoscopic level of description in which the pertinent fields are probability
densities.

Particularly, we have established the kinetic equation for a system in the presence
of entropy barriers. The barriers may be inherent to the intimate structure
of the system or may emerge as a consequence of the elimination procedure of
some coordinates when one tries to simplify its dynamic description.

An illustrative example treated in the literature is the diffusion of a particle
in an region of irregular geometry. The governing equation for the probability
density is known as the Fick-Jacobs equation and has been derived heuristically
by Jacob \cite{Jacob} and directly from the proper coordinate reduction procedure
by Zwanzig \cite{Zwanzig Chem}. In our derivation, the entropic barrier directly
comes out from Boltzmann's entropy, which follows from the proper accounting
of the number of accessible states of the system. In the kinetic equation we
have obtained, the dependency of the diffusion coefficient on the coordinate
follows from the general dependency of the Onsager transport coefficients on
the state variable, according to the rules of nonequilibrium thermodynamics.
This dependency becomes crucial in order to the reduced description accurately
resembles the exact solution. We have proposed a scaling law for the diffusion
coefficient reaching a very good agreement with the exact solution even in the
case when the original Fick-Jacobs equation does not provide a good approximation.
In this way, we have put the validity of a Fick-Jacob-like description in a
more general context.

The theory we have proposed can also be applied to cases in which an energy
barrier is also present. We have obtained the corresponding Fokker-Planck equation
describing its dynamics. Interesting characteristics of this equation are the
coupling of the entropy and energy barriers and the apparent violation of the
Fluctuation Dissipation Theorem, resulting from the elimination of variables.

The theory we have presented is applicable to a wide variety of systems of different
nature. Apart from the diffusion problem in the presence of an entropic barrier
we have discussed in this paper, we can quote protein folding \cite{camacho},
glassy systems \cite{oppenheim}, transport of ions \cite{iones} and macromolecules
\cite{han}-\cite{slater} through membranes or channels, motion of polymers
subjected to rigid constraints \cite{fixman}, protein binding kinetics \cite{binding, ligand},
drug release \cite{drug}, nucleation \cite{shi} or polymer crystallization
\cite{frenkel}, to mention just a few examples in which the presence of entropic
barriers becomes relevant.

There still remain open questions whose answers goes beyond the scope of this
paper. The main pitfalls arise from the coordinate reduction procedure. The
question is how to proceed with the coarsening in more complex situations. For
instance, it will be desirable to analyze the effect of the asymmetries either
intrinsic to the nature of the landscape or due to the presence of external
forces which impose a preferred direction, which occurs in many situations of
practical interest. Another interesting point will be to analyze the behaviour
when the slope diverges, giving rise to regions in which the particle may become
trapped thus breaking the ergodicity of the system. The solutions to these problems
probably demand a more elaborated analysis of the scaling of the diffusion coefficient
which takes into account these effects.

Just proceeding systematically along the lines we have established by means
of our theory one could arrive at a more complete theoretical framework providing
the dynamic description of systems exhibiting entropic barriers.

\section{Acknowledgments}

We want to acknowledge P. Mazur, J.M. Vilar, P. Reimann and G. Schmidg for their
comments and suggestions. D. Reguera wishes to thank Generalitat de Catalunya
for financial support. This work has been partially supported by DGICYT of the
Spanish Government under grant PB98-1258.

\end{document}